
\input phyzzx
\nonstopmode
\sequentialequations
\twelvepoint
\nopubblock
\tolerance=5000
\overfullrule=0pt

\REF\ewpt{A small selection is:
A. Linde, {\it Phys. Lett.} {\bf B70}, 306 (1977);
V. Kuzmin, V. Rubakov and M. Shaposhnikov, {\it ibid} {\bf B155},
36 (1985); M. Dine \etal, {\it ibid} {\bf B257}, 351 (1991);
A. Cohen, D. Kaplan and A. Nelson, {\it ibid} {\bf B245}, 561 (1990);
S. Dimopoulos and L. Susskind, {\it Phys. Rev.} {\bf D18}, 4500 (1978)
P. Arnold and L. McLerran, {\it ibid} {\bf D36}, 581 (1987);
G. W. Anderson and L. J. Hall, {\it ibid} {\bf D45}, 2685 (1992);
M. Carrington, {\it ibid} {\bf D45}, 2933 (1992);
2685 (1992); N. Turok and J. Zadrozny, {\it Phys. Rev. Lett.}
{\bf 65}, 2331 (1990); M. Dine, {\it et al.}, SLAC report
SLAC-PUB-5741 (1992); D. Brahm and S. D. Hsu, Caltech report CALT-68-1705
(1991).}

\REF\hlm{B. I. Halperin, T. Lubensky and S. K. Ma, {\it Phys. Rev.
Lett.} {\bf 32}, 292 (1974).}

\REF\linde{A. Linde, {\it Rep. Prog. Phys.} {\bf 42},
389 (1979).}

\REF\ginsparg{P. Ginsparg, {\it Nucl. Phys.} {\bf B170[FS1]}, 388 (1980).}

\REF\effective{Work in progress.}

\REF\ginzburg{V. Ginzburg, {\it Sov. Phys. Solid State} {\bf 2},
1824 (1961).}

\REF\wilson{K. G. Wilson and M. E. Fisher, {\it Phys. Rev. Lett.}
{\bf 28}, 240 (1972); K. G. Wilson and J. Kogut, {\it Phys. Rep.}
{\bf 12} 75 (1974).}

\REF\ma{An excellent general reference is: S. K. Ma, {\it Modern
Theory of Critical Phenomena}, Benjamin/Cummings, Reading MA, (1976).}

\REF\hl{B. I. Halperin and T. Lubensky, {\it Solid State Comm.} {\bf 14},
997 (1974).}

\REF\expt{C. W. Garland, G. B. Kasting and K. J. Lushington,
{\it Phys. Rev. Lett.} {\bf 43}, 1420 (1979). D. L. Johnson \etal,
{\it Phys. Rev.} {\bf B18}, 4902 (1978).}

\REF\dh{C. Dasgupta and B. I. Halperin, {\it Phys. Rev. Lett.} {\bf 47},
1556 (1981).}

\REF\kane{J. Gunion, \etal, {\it The Higgs Hunter's Guide},
Addison Wesley, Reading MA, (1990).}

\REF\bhzj{E. Brezin, S. Hikami and J. Zinn-Justin, {\it Nucl. Phys.}
{\bf B165}, 528 (1979).}

\REF\cel{T. Cheng, E. Eichten and L.-F. Li, {\it Phys. Rev.}
{\bf D9}, 2259 (1974).}

\REF\hikami{S. Hikami, {\it Prog. Theor. Phys.} {\bf 64}, 1425 (1980).}

\REF\work{M. Alford and J. March-Russell, work in progress.}

\REF\polyakov{A. M. Polyakov, {\it Phys. Lett.} {\bf B59}, 79 (1975);
E. Brezin and J. Zinn-Justin, {\it Phys. Rev. Lett.} {\bf 36}, 691 (1976)
and {\it Phys. Rev.} {\bf B14}, 3110 (1976).}

\REF\amz{D. J. Amit, S.-K. Ma and R. K. P. Zia, {\it Nucl. Phys.}
{\bf B180[FS2]}, 157 (1981).}

\REF\nt{D. R. Nelson, in {\it Phase Transitions and Critical
Phenomena}, Vol.{\bf~7}, ed. C. Domb and J. Lebowitz, Academic
Press, London (1983); D. R. Nelson and J. Toner, {\it Phys. Rev.}
{\bf B24}, 363 (1981).}

\REF\potts{B. Nienhuis, K. Reidel and M. Schick, {\it J. Phys.}
{\bf A13}, L189 (1980), and references therin.}

\REF\wegner{F. Wegner, {\it Z. Phys.} {\bf B35}, 207 (1979).}

\def\fun#1#2{\lower3.6pt\vbox{\baselineskip0pt\lineskip.9pt
  \ialign{$\mathsurround=0pt#1\hfil##\hfil$\crcr#2\crcr\sim\crcr}}}

\def\gap{\mathrel{\mathpalette\fun >}}
\def\order{{\cal O}}
\def\etal{{\it et al.}}

\let\al=\alpha
\let\be=\beta
\let\ga=\gamma
\let\Ga=\Gamma
\let\de=\delta

\let\ep=\epsilon

\let\la=\lambda

\let\del=\nabla
\let\si=\sigma

\let\om=\omega

\let\p=\partial
\let\<=\langle
\let\>=\rangle

\let\ol=\overline

\def\comment#1{ \hbox{Comment suppressed here.} }

\line{\hfill PUPT-92-1328}
\line{\hfill LBL-32540}
\line{\hfill hep-ph/9208215}
\line{\hfill July 1992}
\titlepage
\title{On the Possibility of Second-Order Phase
Transitions in Spontaneously Broken Gauge Theories}
\medskip
\author{John March-Russell}
\smallskip
\centerline{{ Joseph Henry Laboratories, Princeton University}}
\centerline{{ Princeton, NJ 08544\foot{Permanent address.
Research supported by NSF grant NSF-PHY-90-21984. e-mail:
jmr@iassns.bitnet, jmr@puhep1.princeton.edu}}}
\smallskip
\centerline{and}
\smallskip
\centerline{Theoretical Physics Group, Physics Division}
\centerline{Lawrence Berkeley Laboratory}
\centerline{1 Cyclotron Road, Berkeley, CA 94720
\foot{Supported in part by the Director, Office of Energy
Research, Office of High Energy and Nuclear Physics, Division of High
Energy Physics of the U.S. Department of Energy under Contract
DE-AC03-76SF00098.}}
\medskip

In the ``Type-II'' regime, $m_{\rm Higgs}\gap m_{\rm gauge}$,
the finite-temperature phase transition in spontaneously-broken
gauge theories (including the standard model) must be
studied using a renormalization group treatment. Previous
studies within the $(4-\ep)$-expansion suggest a 1st-order
transition in this regime. We use analogies with
experimentally accessible phase transitions in liquid crystals, and
theoretical investigations of superconductor phase transitions to
argue that, in this range, the critical behavior of a large class
of gauge-Higgs-fermion systems changes from 1st to 2nd-order
as a function of Higgs mass. We identify
a set of models which, within the $(2+\ep)$-expansion, possess fixed
points that can describe this 2nd-order behavior. As usual, a definitive
demonstration that the claimed critical behavior occurs (and a reliable
estimate of $m_{\rm Higgs}$ at the tricritical point) will
probably require numerical simulations.
\endpage

\vskip 2in
\centerline{Disclaimer}
\vskip .5in

{\it This document was prepared as an account of work sponsored by the United
States Government.  Neither the United States Government nor any agency
thereof, nor The Regents of the University of California, nor any of their
employees, makes any warranty, express or implied, or assumes any legal
liability or responsibility for the accuracy, completeness, or usefulness
of any information, apparatus, product, or process disclosed, or represents
that its use would not infringe privately owned rights.  Reference herein
to any specific commercial products process, or service by its trade name,
trademark, manufacturer, or otherwise, does not necessarily constitute or
imply its endorsement, recommendation, or favoring by the United States
Government or any agency thereof, or The Regents of the University of
California.  The views and opinions of authors expressed herein do not
necessarily state or reflect those of the United States Government or any
agency thereof of The Regents of the University of California and shall
not be used for advertising or product endorsement purposes.}

\endpage

There has been much recent interest in the physics of the
finite-temperature electroweak phase transition (EWPT) [\ewpt].
The motivation behind this considerable body of work, is the realization
that the observed baryon-number asymmetry of the universe
might be generated at time of the EWPT,
through the anomaly in the conservation laws for the
baryon and lepton number currents.
In relation to this, it is a pleasant fact that classic studies of the
nature of the phase transition in coupled gauge-scalar systems indicate
that they are 1st-order [\hlm,\linde,\ginsparg]. Therefore one of the primary
conditions for a net generation of baryon number, namely that thermal
equilibrium is not maintained, arises naturally in the case of the
standard model due to the supercooling of the false-vacuum
phase and its later decay.
In this letter we will reconsider the question of the order
of the phase transition in a large class of
coupled gauge-scalar-fermion systems (including a version of the
two-doublet standard model) in the
``Type-II'' regime (roughly speaking $m_{\rm Higgs}\gap m_{\rm gauge}$).
We will show that for large enough scalar masses there is reason to
believe that the transition changes over from 1st to 2nd-order, passing
through a tricritical point at some value of the ratio of Higgs
to gauge boson masses.

For Higgs masses in the range we consider our conclusions probably do not
directly impinge on the question of weak scale baryogenesis, since (at
least in many simple extensions of the standard model) even if the
transition were 1st-order, the value of the scalar expectation value
just after the completion of the phase transition is such that the
baryon number asymmetry is ``washed out'' [\ewpt]. Nevertheless,
the mass of the Higgs in our world might well turn out to be in the
Type-II regime, and it is an interesting theoretical question as to the nature
of the transition in that case. The analysis might also have some practical
importance in non-minimal extensions of the standard model and for
other phase transitions, although the we caution the reader that the
critical scalar mass is both model dependent and difficult to estimate.

We start with a review
of the current state of knowledge concerning the order of the phase
transition in gauge-Higgs systems from the perspective of Refs.~[\hlm]
and [\ginsparg].
Consider an abelian gauge field $B_{\mu}$ coupled to a
complex scalar field $\phi$, with action
$$
S=\int d^4r \left( -{1\over 4} F^2 + |D_{\mu}\phi |^2
-m^2|\phi|^2 - {\la\over 4} |\phi|^4 \right),
\eqn\aaction
$$
where $D_{\mu}=\p_{\mu} -ieB_{\mu}$. The study of the finite-temperature
critical behaviour of this theory proceeds by considering
the action in Euclidean space with periodized time of length $\be=1/T$.
The fourier modes of the (periodic) fields $\phi$ and $B$ are then
labelled by a continuous three-dimensional momentum ${\bf k}$ and an
integer $n\in Z$, where $\om_n=2\pi n/\be$ now appears in place of $k^0$.
For a weakly coupled theory (the only case considered in this letter), and
near the transition temperature, $1/\be\sim \mu/g$ (or $\mu/\sqrt{\la}$ if it
is smaller) is then parametrically large compared to the typical mass
scale $\mu$ of the theory, and may be used as an expansion parameter
to isolate the $n=0$ mode [\ginsparg].
This is achieved by integrating out the $n\neq 0$ modes to obtain
an effective three-dimensional theory of the $n=0$ mode alone.
Even though near the transition effective
masses are vanishing, no IR problems arise in this procedure since the
$n\neq 0$ propagators are cutoff by effective ``masses'' $\om_n=2\pi n/\be
\neq 0$.\foot{It is
simple to include fermions in this discussion. Their
antiperiodicity in the time direction implies that $\om_n=(2n+1)\pi/\be$
and thus fermions do not possess zero modes which participate in the
three-dimensional effective action (although they can, of course, affect
the numerical values of its effective coupling constants).}

For the simple case of Eq.~\aaction\  this leads to an
effective three-dimensional theory for the $n=0$ modes of $\phi$
and $B$ of the general form
$$
S_{\rm eff}=\int d^3r \left( {1\over 4} {\bf F}^2 + |{\bf D}\phi |^2
+a|\phi|^2 + {b\over 4} |\phi|^4 + ... \right),
\eqn\freee
$$
where in general all coupling constants have suppressed temperature
dependencies. There is one temperature dependence, however,
that must be kept -- that of the effective mass term $a=a'(T-T_0)$
since it's vanishing at the temperature $T_0$ drives (in mean field
or Landau theory, and ignoring the gauge field) a second order
transition and leads to IR divergences once we take into
account fluctuations. (As we see in a moment this is only the ``transition
temperature'' for a second order phase transition -- we will have to modify our
statements slightly in the first order case.)

The effect of the gauge field on the transition can be qualitatively
understood by formally integrating out the gauge field to
define a free energy (or finite temperature effective action) $F(\phi,T)$
depending only on $\phi$. Since the three dimensional action Eq.~\freee\
is quadratic in $B$ we can evaluate $\< B^2 \>_{\phi}$ near $T_0$ by the
equipartition theorem, leading to
$$
\< B^2 \>_{\phi} \propto T_0 \int d^3k {1\over ({\bf k}^2 + M^2_B(\phi))},
\eqn\expect
$$
where $M_B(\phi)\propto |\phi|$ is the $\phi$ dependent mass
of the gauge boson. Regulating and renormalizing the integral in
Eq.~\expect\ we end up with the finite contribution
$\< B^2 \>_{\phi}\propto -T_0 |\phi|$. We can then substitute this back into
the free energy \freee\ to discover that a term proportional to $|\phi|^3$
with a negative sign has been generated in $F(\phi,T)$ [\hlm].
This, within the framework of mean
field theory, inevitably leads to a 1st-order transition.
However, when fluctuations in $\phi$ are also
considered complications arise.

Define $T_c$ to be the temperature at which
the symmetric ``false'' phase at $\< \phi \> =0$ is degenerate
with the unsymmetric ``true'' phase at $\<\phi \>\neq 0$.
Below $T_c$ the false phase is at best metastable. Define $T_*<T_c$
to be the temperature at which the false phase first becomes
mechanically {\it unstable} (rather than metastable).\foot{$T_*$
is the ``spinodal point''
at which the false phase can first evolve into the true phase
by {\it small amplitude, long wavelength} fluctuations rather than
by the better known (to particle physicists) process of
critical bubble nucleation. Parenthetically, the correctly
defined effective potential for the
false phase should possess an imaginary part in perturbation theory
for low momentum fluctuations corresponding to this decay
process [\effective]. This seems not to be widely appreciated.}
In other words $T_*$ is defined by $d^2 V(\<\phi\>=0,T_*)/d\<\phi\>^2=0$ --
the point of vanishing of the scalar mass around the false phase. To
leading order $T_*$ is equal to $T_0$ above. Because of the IR divergences
that occur in the loop expansion evaluation of the partition function
at $T_*$ there is (for a given size of the effective scalar
couplings in $F(\phi,T)$, and in less than four spatial
dimensions) a range of temperatures around $T_*$ for
which the loop expansion {\it inevitably fails}. This range temperatures
is known as the Ginzburg region [\ginzburg].

This is {\it not} just an academic concern in the case of first order
transition
   s
if $T_c$ is within the Ginzburg region surrounding $T_*$. Generally speaking,
this is the case in the ``Type-II'' parameter range roughly given by
$m_{\rm scalar}\gap m_{\rm gauge}$. For instance, an application
of the Ginzburg criterion to the analogue of $F(\phi,T)$ for the standard model
shows that $T_c$ is within the Ginzburg region for $m_{\rm Higgs}\gap 100$ GeV
(fairly independent of the top quark mass).\foot{There are many equivalent
ways of expressing the ``Ginzburg criterion.'' Probably the simplest in the
field theory context is a direct comparison between the tree and 1-loop
three and four-point functions calculated from the free energy $F(\phi,T)$.}
In this situation the standard
effective potential formalism fails to give any reliable information,
and the only known way to proceed in the study of the critical behavior is
the Wilson-Fisher renormalization group (RG) [\wilson, \ma].

In the Type-II regime consider the coupled
RG equations (within a $(4-\ep)$-expansion) for $\phi$ and $B$
in a slight generalization of the model \aaction\ --
specifically the case of $N$ complex scalar fields.
The RG equations are:
$$
{de^2\over ds}=\be_{e^2}(e^2,\ep)=\ep e^2 - {N\over 24\pi^2}e^4,
\eqn\betae
$$
for the abelian gauge coupling, and
$$
{d\la\over ds}=\be_{\la}(e^2,\la,\ep)=\ep\la-{N+4\over 4\pi^2}\la^2
-{3\over 8\pi^2}e^4 + {3\over 4\pi^2} e^2\la,
\eqn\betala
$$
for the quartic scalar self-coupling. For convenience we have taken $s$
to increase in the IR. We are interested, of course,
in the case $\ep=1$ if we are to describe the critical behaviour
of our model in the physical number of spatial dimensions.
Recall that the RG equations derived in the $\ep$-expansion are only
rigorously true in the limit $\ep\to 0$.
Nevertheless, two decades of experience have shown that
the stable fixed points identified within the expansion lead to a
surprisingly accurate description of many second-order transitions at $d=3$.

For $N\ge 183$ the equations \betae\ and \betala\ possess a stable
fixed point with real couplings, leading to a prediction of a 2nd-order
transition. For a lesser number of scalar fields the only physically
accessible fixed points (i.e. with non-complex values of the couplings) are
the Gaussian and Heisenberg ones, both of
which are unstable with respect to the charge. This lack
of a stable fixed point and the associated runaway of the coupling $e$
was interpreted in Ref.~[\hlm] as the sign of a (weakly) 1st-order transition
even in the Type-II regime. Note that the size of the 1st-order transition
was predicted to be far too small to be experimentally detected. (The
critical region in the high-$T_c$ materials, $|T-T_c|/T_c\sim 10^{-2}$,
is large enough that experiments measuring the non-mean-field critical
properties are feasible.)

In an excellent paper, Ginsparg [\ginsparg] pointed out
that the critical behavior of a finite-temperature four-dimensional
field theory could be described by an effective three-dimensional
theory, as in Eq.~\freee. He also performed an
extension of the work of Ref.~[\hlm] to a large class of non-abelian
gauge-Higgs theories. The result is that,
within the $(4-\ep)$-expansion, theories with an asymptotically free
{\it gauge} coupling constant (not necessarily the entire theory)
have no stable fixed point of the coupled RG equations.
This was interpreted as implying that these
theories also all underwent 1st-order transitions in the Type-II regime.
The physical intuition behind this result is that small,
amplitude fluctuations of the scalar field give the gauge field
a mass, in turn suppressing the gauge field fluctuations which
had tended to disorder the scalar field. Therefore the system
is unstable to a sudden jump in the amplitude of the scalar field.

Unfortunately, the conclusions of Ref.~[\hlm] are known to be
incorrect in this Type-II regime. It is possible to
argue (again within the $(4-\ep)$-expansion) that the critical
properties of the smectic-A to nematic (SAN) phase transition in
liquid crystals are isomorphic to that of the
superconductor [\hl].\foot{Note that the free-energy describing this
transition is of essentially the same form as the
superconductor -- the director field in the smectic-A phase playing
the role of the vector potential. There are differences though --
especially the spatially anisotropic nature of the smectic-A phase
which makes comparison with critical theories delicate.}
Thus the prediction is that this phase transition should similarly
be weakly 1st-order. The liquid-crystal case differs from that of the
superconductor in that the size of the latent heat was such that it
could easily be detected. Experimentally, however, the SAN transition is
observed to be second order with approximately XY
exponents [\expt]. (The situation is complicated by the evidence of
``anisotropic scaling.'')

Stimulated by these findings, the Type-II superconductor transition in
three-dimensions has been reconsidered [\dh]. By starting with
a lattice model, and applying duality arguments, the partition function
of the superconductor can be mapped onto that of a set of directed strings
with repulsive contact interactions. With the aid of monte carlo
simulations, this system was then studied and shown to lead to a 2nd-order
transition with XY-model exponents (but with inverted asymmetry of
the amplitudes of the singular terms with respect to
temperature). Therefore a new fixed point of the RG
equations must exist at $\ep=1$ {\it which we cannot see
by analytically continuing away from} $d=4$.

Indeed, as we will argue in detail below, fluctuations which drive
the critical behaviour of a system 2nd-order get stronger as we
approach two dimensions. As we approach four dimensions 1st-order
behaviour is favored. Thus, roughly speaking, the most reliable
predictions of expansions in $(4-\ep)$-dimensions are that of 2nd-order
transitions. The obvious consequence for non-abelian theories is that
care must be taken in identifying the lack of a stable fixed-point
in the expansion away from four-dimensions with a 1st-order
transition.

If the theories that we are considering possess a 2nd-order transition
then there must exist a fixed point of the RG equations. If the $(4-\ep)$
expansion fails to find this fixed point, then how are we to proceed?

To be specific, what can we say about the critical properties of
the following three-dimensional Lagrangian describing $N$ complex
$p$-vectors coupled to an $SU(p)\times U(1)$ gauge-theory?
$$
L=|{\bf D}\phi^a|^2 + {1\over 4}{\bf G}^2 + {1\over 4}{\bf F}^2
+ \la ({\ol\phi}^a\cdot\phi^a)({\ol\phi}^b\cdot\phi^b) +
\ga ({\ol\phi}^a\cdot\phi^b)({\ol\phi}^b\cdot\phi^a)  + {\rm mass~terms},
\eqn\supaction
$$
where $D_{\mu}=\p_{\mu} - ieB_{\mu} -igA_{\mu}$, with $A$ and $B$
the $SU(p)$ and $U(1)$ gauge fields respectively ($G$ and $F$ are
the associated field strengths). Here $a,b=1,...,N$, repeated
indices are summed, and ${\ol\phi}\cdot\phi$ denotes the inner
product among $p$-vectors. An extra $SU(N)$ global symmetry has been
imposed on the scalar sector so that tractable RG equations result.
This is an obvious generalization of the standard-model.\foot{In the
$SU(2)\times U(1)$ two-doublet case ($p=2$, $N=2$), Eq.~\supaction\
corresponds to $\la+\ga=\la_3$, $\ga=-\la_4/2$, $\la_1=\la_2=\la_5=\la_6=0$
and $4\la_3 v^2=m^2$ with $v^2_1=v^2_2=v^2$, in the notation
of [\kane]  (before finite-$T$ effects are taken into account).
Here the zero-temperature mass term is $-m^2{\ol\phi}^a\cdot\phi^a$.}

Consider the apparently unrelated gauged non-linear $\si$-model
in two dimensions [\bhzj], defined by the action
$$
S_\si={1\over t}\int d^2x \left( |\p_\mu \phi^a_i|^2
+ {1\over2}\left(({\ol\phi^a_i}\p_\mu \phi^a_j)({\ol\phi^b_j}\p_\mu
\phi^b_i) + h.c.\right)\right),
\eqn\simodel
$$
with the additional constraint ${\ol\phi^a_i} \phi^a_j = \de_{ij}$.
Here $i,j=1,...,p$ and $a,b=1,...,N$, so that this $\si$-model is that
of the complex grassmann manifold $U(N)/U(p)\times U(N-p)$.
Note that in the action Eq.~\simodel, we have already
integrated out a $U(p)$ gauge field that originally appeared
in covariant derivatives of $\phi$. This leads to the second term
in Eq.~\simodel. (The elimination of the gauge field is exact, since,
by definition, it possesses no kinetic term.)
The beta-function for the single coupling $t$ of this model
($t$ is a dimensionless parameter proportional to the temperature)
up to three-loops and in $(2+\ep)$-dimensions is [\bhzj]:
$$
\be_t(t)=-\ep t + N t^2 + 2(p(N-p) +1)t^3 + \order(t^4).
\eqn\sibeta
$$
This theory is, of course, asymptotically free in the coupling $t$ at
$d=2$ (we have again defined $s$ to increase in the IR). In the language of
statistical-mechanics, this translates in the statement that we have
an IR-unstable fixed point at a critical temperature $t_c=\ep
-2\ep^2\big(p(N-p) + 1\big)/N + \order(\ep^3)$,
resulting in a 2nd-order transition.\foot{Recall that
with respect to temperature a fixed point
describing a 2nd-order transition is {\it unstable}, since dilation of
the coordinates decreases the effective correlation length, and
therefore, increases the reduced temperature $|T-T_c|/T_c$.}
The correlation length exponent $\nu$ (defined by $\xi(t)\sim
|t-t_c|^{-\nu}$) is simply related to $\be_t$;
$$
\nu^{-1}=-d\be_t/dt|_{t_c}=\ep + 2\big(p(N-p)+1\big)\ep^2/N^2
+\order(\ep^3),
\eqn\sinu
$$
which in the large $N$ limit reduces to
$$
\nu^{-1}=\ep+2p\ep^2/N.
\eqn\sinuN
$$
(It is also simple to calculate other exponents after
applying the well known scaling laws [\ma] to $\nu$ and the
anomalous dimension of the operator ${\ol\phi^a}\cdot \phi^b
-(p/N)\de_{ab}$ [\bhzj].)

Returning to the action Eq.~\supaction\  we may also consider
its beta-functions, but
in $(4-\ep)$-dimensions. Standard calculations show that they are
Eq.~\betae\ with the replacement $N\to pN$ supplemented by
$\be_g=\ep g^2 + {g^4\over 8\pi^2}\left({11p\over 3}-{N\over 6}\right)$,
and some rather complicated expressions for $\be_\la$ and $\be_\ga$
[\cel]. These equations have a stable fixed point
in the large $N$ limit (at finite $p$) with an associated set of
critical exponents. For instance the exponent $\nu$ is given by
$$
\nu^{-1}=2-\ep + 48p\ep/N +\order(1/N^2).
\eqn\nuexptfour
$$
We may also study the RG equations for the action \supaction\ directly
within the $1/N$ expansion [\ma] and for arbitrary dimension $2<d<4$.
For instance we now find [\hikami]
$$
\nu={1\over d-2}\left(1+{2(d^2-d)\sin(d\pi/2)\Ga(d-1)p\over
\pi N \Ga^2(d/2)}\right).
\eqn\nuN
$$
Along with the other exponents this can be expanded near {\it both} four
and two dimensions: for $\ep=(4-d)$ we recover Eq.~\nuexptfour, while
for $\ep=(d-2)$ we obtain Eq.~\sinuN\ [\hikami].
Similar relations hold between the values of the other critical
exponents calculated (where they are well defined), in the large
$N$ limit, for, respectively, the $\si$-model near two dimensions,
and the original $SU(p)\times U(1)$ gauge theory near four dimensions.
The interpretation of these relations we wish to emphasize
is that, in the large-$N$ limit,
both the $(4-\ep)$ and $(2+\ep)$-expansions are in fact describing
the {\it same} fixed point, albeit from differing perspectives.

What about the nature of the phase
transition in the system \supaction\ for small $N$? Define the
continuous function $d(N)$ as the number of (spatial) dimensions
at which, for a given $N$, the transition switches over from 1st
to 2nd-order ($d(N)$ is clearly model-dependent). The curve $d(N)$
in $(N,d)$-space is then a line of singularities through which it
is impossible to analytically continue \FIG\dndiag{The character of
the critical behavior of the $SU(p)\times U(1)$ gauge theory coupled
to $N$ $p$-vectors, in $d$-dimensions, changes from 1st to 2nd-order
as we cross the curve $d(N)$. Expansions in $\ep=(4-d)$ inevitably
break down as $d(N)$ is approached. The 2nd-order transitions
occuring along the hatched line at $d=3$, are only accessible via
$(2+\ep)$-expansions away from the gauged non-linear sigma-model
formulated in two-dimensions.} (see Fig.~\dndiag).
The results of the $(4-\ep)$-expansion tell us that at some
critical value $N_c$, $d(N)$ drops infinitesimally below four
dimensions (considering $N$ as a continuous parameter which we
decrease away from large values).
The important point is then the following: If we assume the existence
of a fixed point at $d=3$, even below $N_c$, then as we smoothly change
$N$ from just above, to just below $N_c$, we expect, by continuity,
that the properties of the fixed point at $d=3$ do not greatly alter.
So if we have a description of the fixed point that is smooth through
$N_c$, then this should give an at least qualitatively correct
description of its properties into the region $N<N_c$.
But the RG analysis of the system Eq.~\simodel\ within the
$(2+\ep)$-expansion is exactly such a description!

We therefore propose that the IR behavior of the $SU(p)\times U(1)$
gauge model in three dimensions and for $N<N_c$ is described by
the gauged non-linear $\si$-model of Eq.~\simodel. In particular
the correlation exponent $\nu$ of the
$SU(p)\times U(1)$ gauge theory in three
spatial dimensions is approximated by the expression Eq.~\sinu\ at
$\ep=1$. Furthermore, using the Josephson scaling law $\al=2-d\nu$ [\ma]
we may extract a prediction for the specific heat exponent $\al$.
For the interesting case of $N=p=2$ this gives $\nu=2/3$ and $\al=0$.
It it also possible to extract other exponents from the formulae of
Ref.~[\bhzj], although we will not do so here. We note that in cases
where both $(4-\ep)$ and $(2+\ep)$ expansions predict 2nd-order
phase transitions, the expansion away from two dimensions is generally
less quantitatively reliable, and quite sophisticated techniques are
required to extract accurate exponent values at $d=3$. Thus Eq.~\sinu\
and the associated value of $\al$ should be taken as a rough
guide only. We also caution the
reader that we have assumed $d(N)$ is a monotonically decreasing
function of $N$ (as $N$ is decreased), and that $d(N)\ge 3$ for
values of $(N,p)$ as low as $(2,2)$. Although we have
the analogies with the superconducting and liquid-crystal systems to
support these assumptions it is possible to be misled, so that
numerical simulations are really required to settle the issue
definitively. These simulations would also enable us to find the
value of the ratio of Higgs to gauge boson masses at the
tricritical point, at which the system switches over from 1st to
second order behavior. For the interesting case of small $N$, this
ratio is likely to be hard to extract analytically.

To better understand the physics behind the arguments of the
preceeding paragraphs it is useful to consider a simpler, purely
scalar, system [\amz,\nt].
Let $\phi$ be an $n$-component complex scalar field with
three-dimensional free-energy given by
$$
F=\int d^3x\left( |{\bf \del}\phi|^2  - a |\phi|^2 +  b |\phi|^4 + ...\right).
\eqn\scalara
$$
Consider the limit in which $b\to\infty$ with $a/2b$ fixed. In this limit
the value of the field $\phi$ becomes very well localized around its
vacuum expectation value $\<|\phi|^2\>=a/2b$. However, the $(n-1)$
Goldstone modes are free to fluctuate, and dominate the IR behavior.
Indeed, formally integrating out the ``amplitude fluctuations'' in this limit
we arrive at the free-energy of the non-linear $O(n)$ sigma-model,
$$
F=\int d^3x ({\bf\del}\phi)^2,
\eqn\scalarnlsm
$$
with the constraint $|\phi|^2=a/2b$. In this free-energy only the ``phase
fluctuations'' appear, and the associated system is known to undergo
a 2nd-order transition near $d=2$ [\polyakov].
Note that {\it for a fixed amplitude} of $\phi$,
the addition of a cubic term to \scalara\ (so that a 1st-order transition is
naively predicted for this model), has, almost by definition,
no effect -- we still arrive at Eq.~\scalarnlsm, with its 2nd-order
transition. Of course, for any finite $b$, we cannot just
throw away the amplitude fluctuations -- the correct procedure is to
consider the RG flow of the couplings of the (higher-dimension)
operators that now appear in the action \scalarnlsm\ when the amplitude
of $\phi$ is allowed to fluctuate. A very important result of this
procedure for our discussion is that, near two-dimensions,
all these higher-dimension operators are {\it irrelevant} (in the
Wilson-Fisher sense) [\amz,\nt] and so do not affect the order
of the transition. (To our knowledge, no
explicit proof of the analogous statement for gauged non-linear
sigma-models has been presented. Nevertheless we expect that
similar statements continue to hold [\work].)
One way to think about such phase-fluctuation-driven phase
transitions is that near $d=2$ the true transition temperature
is driven well below its mean-field value $T_c^{({\rm mf})}$
(recall that $T_c\propto\ep$ [\polyakov]),
so that the effective mass term of the amplitude fluctuations,
$a\propto (T_c^{({\rm mf})}-T)$, is actually quite
large at the transition.
Further note that generally speaking, the effect of these
phase-fluctuations {\it increases}
as we move to more complex non-abelian systems. This gives us some
reason to hope that the analogies to superconductor and liquid-crystal
systems (which are, of course, abelian) may not be misleading.

We can heuristically include gauge fields $A$ in this discussion by
adding to Eq.~\scalara\ a gauge kinetic term $g^{-2}F^2$,
and changing derivatives to covariant derivatives. Now, if in addition
to considering fixed amplitude $\phi$-fields, we also take the
formal limit $g\to\infty$, then we can integrate out $A$ leading
to an action of the form Eq.~\simodel. Despite the brutal nature
of these manipulations, the large-$N$ expansion and analogies
presented above, indicate that the resulting theory may give a
good description of the (static) far-IR behavior of the
original theory in three dimensions. This procedure also generalizes
to models other than the $SU(p)\times U(1)$ gauge system considered
above.

A similar situation occurs for the $q$-state Potts model for
$q=3,4$ [\potts]. For
$q\ge 3$ such models possess a cubic invariant and Landau theory predicts
a 1st-order phase transition. Recall that the lower-critical dimension
for systems with a discrete symmetry is $d=1$, so that as we approach
$d=1$ from above, fluctuations drive the transition temperature to
zero. Studies of the RG in $(1+\ep)$-dimensions show that all Potts
models possess a 2nd-order transition near $d=1$. The interesting
point is that it is rigorously known that the three-state Potts
model is 2nd-order upto $d=2$ (the four-state Potts model also has a
continuous transition at $d=2$, although the situation is more
complicated due to the merging of the critical and tricritical
fixed points). This is despite the fact that a
$(6-\ep)$-expansion (analogous to the $(4-\ep)$-expansion we considered
previously) finds no stable fixed point. The interpretation is again
that ``phase'' fluctuations have suppressed amplitude fluctuations.
Our proposal for the critical behaviour of the gauged systems
is that they are similarly ``phase-fluctuation-driven'' 2nd-order
transitions, but now in three dimensions.

We wish to emphasize that many models other than the
$SU(p)\times U(1)$ example of Eq.~\supaction\
may be analyzed in the framework described above, and that
this analysis leads to similar results.
(Of particular interest are the $O(p)$
gauge theories -- and related non-linear sigma models --
discussed in Refs.~[\bhzj] and [\hikami].)
Namely, within a $(2+\ep)$-dimensional
renormalization-group analysis, and in the Type-II regime,
{\it 2nd-order} transitions are
predicted. Work is in progress concerning these
theories, as well as predictions for the additional critical exponents,
and an analysis of the RG flows of the higher-dimension operators,
of the $SU(p)\times U(1)$ systems [\work].

Finally, we mention that it might be possible to
extend our predictions of the critical behavior of the $SU(p)\times U(1)$
systems all the way down to $p=2$ and $N=1$ -- the minimal
standard model. Although, for these
values, the action Eq.~\simodel\ no longer makes sense,
the critical exponents (considered as analytic functions of $N$ and $p$)
might still be qualitatively correct. This is similar to considerations
made in the study of Anderson localization [\wegner], and $0$-component
spins applied to random walks [\ma]. It is also possible that our results
have some relevance to the difficulties encountered in the study of the
EWPT in the Type-I regime [\ewpt]. For instance, the proximity of a
2nd-order phase transition could result in an anomalously weak 1st-order
transition in this regime.

I wish to thank Mark Alford and Frank Wilczek for stimulating
conversations, Bert Halperin, Tom Lubensky, and David Nelson
for conversations and discussions of their work, and Paul Ginsparg
for helpful comments on a draft version. I am also very
grateful to the LBL theory group for their hospitality during
the course of this investigation.

\par

\refout
\figout

\end